\documentclass[twocolumn,prl]{revtex4}
\usepackage{epsfig,amsmath,amssymb,graphics,color,calc}

\begin{document}

\newcommand{\bq}{\ensuremath{{\bf q}}}
\renewcommand{\cal}{\ensuremath{\mathcal}}
\newcommand{\bqp}{\ensuremath{{\bf q'}}}
\newcommand{\bbq}{\ensuremath{{\bf Q}}} 
\newcommand{\bp}{\ensuremath{{\bf p}}}
\newcommand{\bpp}{\ensuremath{{\bf p'}}}
\newcommand{\bk}{\ensuremath{{\bf k}}}
\newcommand{\bx}{\ensuremath{{\bf x}}}
\newcommand{\bxp}{\ensuremath{{\bf x'}}}
\newcommand{\by}{\ensuremath{{\bf y}}}
\newcommand{\byp}{\ensuremath{{\bf y'}}}
\newcommand{\bxpp}{\ensuremath{{\bf x''}}}
\newcommand{\rmd}{\ensuremath{{\rm d}}}
\newcommand{\intk}{\ensuremath{{\int \frac{d^3\bk}{(2\pi)^3}}}}
\newcommand{\intq}{\ensuremath{{\int \frac{d^3\bq}{(2\pi)^3}}}}
\newcommand{\intqp}{\ensuremath{{\int \frac{d^3\bqp}{(2\pi)^3}}}}
\newcommand{\intp}{\ensuremath{{\int \frac{d^3\bp}{(2\pi)^3}}}}
\newcommand{\intpp}{\ensuremath{{\int \frac{d^3\bpp}{(2\pi)^3}}}}
\newcommand{\intx}{\ensuremath{{\int d^3\bx}}}
\newcommand{\intxp}{\ensuremath{{\int d^3\bx'}}}
\newcommand{\intxpp}{\ensuremath{{\int d^3\bx''}}}
\newcommand{\drho}{\ensuremath{{\delta\rho}}}
\newcommand{\rhoh}{\ensuremath{{\widehat{\rho}}}}
\newcommand{\fh}{\ensuremath{{\widehat{f}}}}
\newcommand{\phih}{\ensuremath{{\widehat{\phi}}}}
\newcommand{\thetah}{\ensuremath{{\widehat{\theta}}}}
\newcommand{\etah}{\ensuremath{{\widehat{\eta}}}}
\newcommand{\0}{\ensuremath{{(\bk,\omega)}}}
\newcommand{\x}{\ensuremath{{(\bx,t)}}}
\newcommand{\xp}{\ensuremath{{(\bx',t)}}}
\newcommand{\xtp}{\ensuremath{{(\bx',t')}}}
\newcommand{\xtpp}{\ensuremath{{(\bx'',t')}}}
\newcommand{\xttpp}{\ensuremath{{(\bx'',t'')}}}
\newcommand{\xtpn}{\ensuremath{{(\bx',-t')}}}
\newcommand{\xtppn}{\ensuremath{{(\bx'',-t')}}}
\newcommand{\xn}{\ensuremath{{(\bx,-t)}}}
\newcommand{\xpn}{\ensuremath{{(\bx',-t)}}}
\newcommand{\xppn}{\ensuremath{{(\bx',-t)}}}
\newcommand{\xpp}{\ensuremath{{(\bx'',t)}}}
\newcommand{\xxp}{\ensuremath{{(\bx,t;\bx',t')}}}
\newcommand{\Crr}{\ensuremath{{C_{\rho\rho}}}}

\newcommand{\Crf}{\ensuremath{{C_{\rho f}}}}
\newcommand{\Crt}{\ensuremath{{C_{\rho\theta}}}}
\newcommand{\Cff}{\ensuremath{{C_{ff}}}}
\newcommand{\Cffh}{\ensuremath{{C_{f\fh}}}}
\newcommand{\Ct}{\ensuremath{{\dot{C}}}}
\newcommand{\Ctt}{\ensuremath{{\ddot{C}}}}
\newcommand{\Crrp}{\ensuremath{{\dot{C}_{\rho\rho}}}}
\newcommand{\Crfp}{\ensuremath{{\dot{C}_{\rho f}}}}
\newcommand{\Crtp}{\ensuremath{{\dot{C}_{\rho\theta}}}}
\newcommand{\Cffp}{\ensuremath{{\dot{C}_{ff}}}}
\newcommand{\Crrpp}{\ensuremath{{\ddot{C}_{\rho\rho}}}}
\newcommand{\thetab}{\ensuremath{{\overline{\theta}}}}
\newcommand \be  {\begin{equation}}
\newcommand \bea {\begin{eqnarray} \nonumber }
\newcommand \ee  {\end{equation}}
\newcommand \eea {\end{eqnarray}}
%\nofiles
%%\begin{frontmatter}

\title{On the top eigenvalue of heavy-tailed random matrices}
\author{Giulio Biroli$^{1,3}$, Jean-Philippe Bouchaud$^{2,3}$, Marc Potters$^{3}$}
%\email{bouchau@spec.saclay.cea.fr}
\affiliation{
$^1$ Service de Physique Th{\'e}orique,
Orme des Merisiers -- CEA Saclay, 91191 Gif sur Yvette Cedex, France.}
\affiliation{
$^2$ Service de Physique de l'{\'E}tat Condens{\'e},
Orme des Merisiers -- CEA Saclay, 91191 Gif sur Yvette Cedex, France.}
\affiliation{
$^3$ Science \& Finance, Capital Fund Management, 6 Bd
Haussmann, 75009 Paris, France.}

\begin{abstract}
We study the statistics of the largest eigenvalue $\lambda_{\max}$ of
$N \times N$ random matrices with {\sc iid} entries of variance $1/N$, 
but with power-law tails $P(M_{ij})\sim |M_{ij}|^{-1-\mu}$. When $\mu >
4$, $\lambda_{\max}$ converges to $2$ with Tracy-Widom fluctuations of
order $N^{-2/3}$, but with large finite $N$ corrections. When $\mu < 4$, 
$\lambda_{\max}$ is of order
$N^{2/\mu-1/2}$ and is governed by Fr\'echet statistics. The marginal
case $\mu=4$ provides a new class of limiting distribution that we
compute explicitly. We extend these results to sample covariance
matrices, and show that extreme events may cause the largest
eigenvalue to significantly exceed the Mar\v{c}enko-Pastur edge.
\end{abstract}

\maketitle

One of the most exciting recent result in mathematical physics is the Tracy-Widom distribution of the 
top eigenvalue of large random matrices \cite{TW}. In itself, this result is remarkable since it constitutes one 
of the rare exactly soluble case in extreme value statistics for strongly correlated random variables (the
eigenvalues of a random matrix), generalizing in a non trivial way the well known Gumbel-Fisher-Tippett, Weibull and
Fr\'echet cases \cite{Galambos}. But the truly amazing circumstance is that the very same distribution appears 
in a host of physically important problems \cite{Spohn}: crystal shapes, exclusion processes \cite{Satya}, sequence matching, 
directed polymers in random media, etc. The last case can in fact be considered, thanks to the mapping onto
the Tracy-Widom problem, as an exactly soluble disordered system in finite dimensions, for which not only the scaling
exponents but the full distribution of the ground state energy can be completely characterized \cite{Johansson}. 

As for many limit theorems, the Tracy-Widom result is in fact expected to hold for a broad class of random matrices. 
The precise 
characterisation of this class, as well as the extension of the Tracy-Widom result for other classes, is a subject of 
intense activity \cite{Sosh1,BBAP}. It is already known that for symmetric $N \times N$ matrices ${\bf M}$ with {\sc iid} entries $M_{ij}$ of variance $1/N$,  
such that all moments are 
finite, the Tracy-Widom result holds asymptotically \cite{Sosh1}. The case where the distribution of entries decays as a power-law 
$\sim |M_{ij}|^{-1-\mu}$ (possibly multiplying a slow function) is expected to fall in a different universality class, at 
least when $\mu$ is small enough. In the case $\mu < 2$ where the variance of entries diverge, it is known that even the 
eigenvalue spectrum $\rho(\lambda)$ of ${\bf M}$ is no longer 
the Wigner semi-circle but itself acquires power-law tail 
$\rho(\lambda) \sim |\lambda|^{-1-\mu}$, bequeathed from the tails of the matrix entries \cite{CB,Burda}. Correspondingly, the largest 
eigenvalues are described by Fr\'echet statistics \cite{Sosh2}. What happens when $\mu$ is in the range $]2, +\infty)$, such that 
the eigenvalue spectrum $\rho(\lambda)$ still converges \cite{CB}, for large $N$, to the Wigner semi-circle? The aim of this letter is 
to discuss this problem in details. We find that as soon as $\mu > 4$, the Tracy-Widom result holds asymptotically, albeit 
with large finite size corrections that we compute. For $\mu < 4$, the largest eigenvalues are still ruled 
by Fr\'echet statistics. The marginal case $\mu=4$ provides a new class of limiting distribution that we compute 
explicitly. We then extend these results to the case of sample covariance Wishart matrices, for which power-law tailed elements 
are extremely common, for example in financial applications \cite{book}. Finally, the relation with directed polymers in the presence 
of power-law disorder is shortly addressed. 

We start by considering real symmetric matrices with {\sc iid} elements of variance equal to $1/N$, and such that the 
distribution has a tail decaying as: 
\be\label{dist}
P(M_{ij}) \simeq \frac{\mu (A N^{-1/2})^\mu}{|M_{ij}|^{1+\mu}},
\ee
where the tail amplitude 
insures that $M_{ij}$'s are of order $A N^{-1/2}$. As soon as $\mu > 2$, the density of eigenvalues converges 
to the Wigner semi-circle on the interval $\lambda \in [-2,2]$, meaning that the probability to find an eigenvalue 
beyond $2$ goes to zero when $N \to \infty$. However, this does not necessarily mean that the largest eigenvalue 
tends to $2$ -- we will see below that this is only true when $\mu > 4$. In order to understand the statistics of 
the largest eigenvalues, we need first to study the following auxiliary problem. Consider an $N \times N$ 
random matrix $\widehat{\bf M}$ with {\sc iid} elements $\widehat M_{ij} \sim N^{-1/2}$, such that its eigenvalue spectrum is, 
for large $N$, the Wigner semi-circle. Now, we perturb this matrix by adding a certain amount $S$ to a given 
pair of matrix elements, say $(\alpha,\beta)$: $\widehat M_{\alpha \beta} \to \widehat M_{\alpha \beta} + S$ and
$\widehat M_{\beta \alpha} \to \widehat M_{\beta \alpha} + S$. What can one say about the spectrum of this new matrix? There
are several ways to solve this problem: self-consistent perturbation theory (that we use 
below), free convolution methods \cite{Verdu} or the replica method; the last two methods in principle require some specific properties of matrix $\widehat{\bf M}$, 
for example that $\widehat{\bf M}$ has Gaussian entries. However, the three methods give the same results for large $N$, 
as can be understood from general diagrammatic considerations (see e.g. \cite{Zee}). Self-consistent perturbation theory is rather 
straightforward and can be easily extended to other cases, such as Wishart matrices (see below). We write down the
eigenvalue equations as:
\begin{equation}\label{e3}
\sum_{j \neq \alpha,\beta}\widehat M_{i,j}v_{j}+\widehat M_{i,\alpha}v_{\alpha}+\widehat M_{i,\beta}v_{\beta} =\lambda v_{i};
\quad i \neq \alpha,\beta
\end{equation}
while for $i=\alpha$, neglecting $\widehat M_{\alpha,\beta}$ compared to $S$:
\begin{equation}\label{e1}
\sum_{j \neq \alpha,\beta}\widehat M_{\alpha,j}v_{j}+ S v_\beta+ \widehat M_{\alpha,\alpha}v_{\alpha}=\lambda v_\alpha,
\end{equation}
and similarly for $i=\beta$. We look for a special solution such that $v_\alpha=v_\beta=v^*$ is of order unity, whereas 
all other $v_i$'s are of order $N^{-1/2}$. We assume (as will be self-consistently checked) that  in the large $N$ limit 
the terms $\sum_{i \neq \alpha,\beta}\widehat M_{\alpha,i}v_{i}$ and $\sum_{j \neq \alpha,\beta}
\widehat M_{\beta,j}v_{j}$ both 
converge to $Kv^*$, where $K$ is a constant to be determined. As a consequence, from Eq. (\ref{e1}), $\lambda=S+K$ up to
small corrections. One can now solve equation (\ref{e3}) to obtain:
\be\label{e4}
v_{i}=\sum_{\ell,j=1}^{N-2}\frac{1}{S+K-\eta_\ell}w_{i}^\ell w_{j}^\ell [\widehat M_{j,\alpha}+\widehat M_{j,\beta}]v^*,
\ee
where $\eta_\ell$ and $w_i^\ell$ are the eigenvalues and the eigenvectors of the $N-2\times N-2$ matrix obtained 
from $\widehat M$ removing the rows and the columns $\alpha$ and $\beta$. Using this expression, we can compute 
$\sum_{i \neq \alpha,\beta} \widehat M_{\alpha,i} v_{i}$. Up to terms negligible in the large $N$ limit one finds: 
\be 
\sum_{i \neq \alpha,\beta} \widehat M_{\alpha,i} v_{i} \approx \int {\rm d \eta}\, \rho_W(\eta) \frac{v^*}{K+S-\eta}
\ee
where $\rho_W(\eta)$ is, by assumption, the Wigner semicircle. Performing the integral over $\eta$, 
the above self-consistency assumption finally leads to $K+S \mp \sqrt{(K+S)^2-4}=2K$. This equation for $K$
only has a solution when $|S| \ge 1$, in which case $K=1/S$ and the corresponding eigenvalue of the perturbed
matrix is $\lambda=S+1/S$ with $|\lambda| \geq 2$, which is therefore expelled from the Wigner sea (see \cite{BBAP} for a similar
mechanism in the case of sample covariance matrices, and \cite{Peche} for Hermitian random matrices). Note that these 
eigenvalues come in pairs, with $\lambda=-S-1/S$, corresponding to $v_\alpha=-v_\beta=v^*$. When $|S| <1$, one the other hand,
no such eigenvalue exist, our assumption that there exists a localised eigenvector sensitive to the presence of $S$
breaks down, and the edge of the spectrum remains $\lambda_{\max}=2$ in this case. One can in fact compute $v^*$ 
and characterize completely the corresponding localised eigenvector. Using Eq. (\ref{e4}) and imposing the normalisation
condition $2v^{*2}+\sum_{i\neq \alpha,\beta}v_{i}^2=1$ one finds: 
\be \label{vstar}
v^{*2}=\frac{1}{2\left[1+\int {\rm d \eta}\, \rho_W(\eta) \frac{1}{(S+K-\eta)^2} \right]} = \frac{1}{2}
\left(1-S^{-2}\right),
\ee
showing that for large $S$, as expected, the eigenvector completely localises on $\alpha$ and $\beta$, whereas when
$|S| \to 1$, it ``dissolves'' over all sites -- for $|S| <1$ the perturbation is not strong enough to induce condensation.
Eq. (\ref{vstar}) enables one to compute various participation ratios of the eigenvectors, for example $w_4=\sum_i v_i^4
=(1-S^{-2})^2/2$. 

The above computation shows that adding any entry strictly less than unity (in absolute value) to a Wigner matrix
does not affect the statistics of its largest eigenvalue. There is in fact a stronger theorem, due to S. P\'ech\'e 
\cite{Peche}, showing that the addition of any matrix of rank $< \epsilon N$, 
$\epsilon \to 0$, with its largest eigenvalue $\Lambda$ less that unity, 
leaves 
unchanged  the statistics of the largest eigenvalue of a random Hermitian matrix. The mechanism leading to such a result 
is very similar to the one above; the largest eigenvalue of the resulting matrix is $\Lambda+\Lambda^{-1}$ when 
$\Lambda > 1$, and $2$ otherwise. 

\begin{figure}
\begin{center}
\psfig{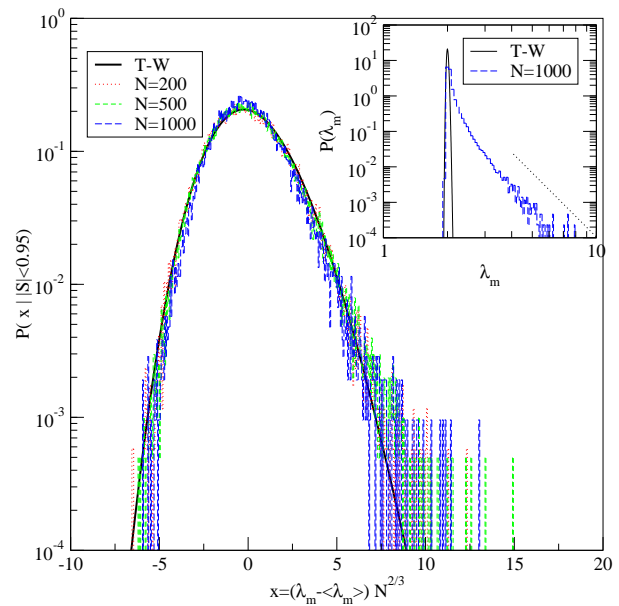} 
\end{center}
\caption{Histogram of $\lambda_{\max}$ conditioned on
$|S|<0.95$ for $\mu=5$ for $N=200,500,1000$ each eigenvalue has been
shifted by the empirical mean and scaled by $N^{2/3}$, for comparison
a GOE Tracy-Widom distribution of zero mean and variance adjusted to
match $N=500$ data is also shown (data obtained from
\cite{Prahofer}). Similar agreement with Tracy-Widom and scaling in
$N^{2/3}$ is obtained for any value of $\mu$ when conditioned on
$|S|<0.95$. Note that for the parameters chosen here, the probability of
$|S|>1$ is still quite large (75.2\%) at $N=1000$. Even for such
large values of $N$, the unconditional distribution of
$\lambda_{\max}$ has a marked power-law tail of index $\mu$ (dotted line) and is very
different from the asymptotic Tracy-Widom distribution
(inset).}
\end{figure}

We can now come back to our initial problem, and define the matrix $\widehat {\bf M}$ by removing all elements of 
$\bf M$ that are (in absolute value) larger than $C N^{-1/2}$, where $C$ is finite, but as large as we wish. 
It is clear that all the moments of $\widehat {\bf M}$ are now finite; therefore the largest eigenvalue of 
$\widehat {\bf M}$ has a Tracy-Widom distribution of width $N^{-2/3}$ around $\lambda_{\max}=2$ \cite{Sosh1,TW}. 
The number of `large' entries that we have removed is, using Eq. (\ref{dist}), $N^2 \epsilon$ with 
$\epsilon=(A/C)^\mu \ll 1$. Now, we should add back the entries that we have left out, starting by all those 
between $C N^{-1/2}$ and $1^-$. Naively, each one of them leaves $\lambda_{\max}$ unchanged: one can dress 
$\widehat {\bf M}$ with all entries less that $1$ and still keep the largest eigenvalue Tracy-Widom. If the number
of such elements was $N \epsilon$ (with $\epsilon \to 0$), the P\'ech\'e theorem would insure that this is true. 
Unfortunately, this number is rather $N^2 \epsilon$, but all added entries are {\sc iid} and randomly scattered over 
the matrix, and the Wigner semi-circle is preserved at each step. It is thus natural to conjecture that provided all
these entries are strictly below unity, the largest eigenvalue remains Tracy-Widom. We have checked this numerically 
for different values of $\mu$ (see Fig.\ 1).

We are now left with entries $|M_{ij}| > 1$. From Eq. (\ref{dist}),
their number is $N^2 \int_1^\infty P(M_{ij}) dM_{ij} = A^\mu
N^{2-\mu/2}$. In the case $\mu > 4$, it is clear that this number
tends to zero when $N \to \infty$. With probability close to unity for large $N$, no
entry is larger than one, in which case
the largest eigenvalue is Tracy-Widom. With small probability, the
largest element $S$ of ${\bf M}$ exceeds one; its distribution is
$A^\mu N^{2-\mu/2}/|S|^{1+\mu}$ and the corresponding largest
eigenvalue, using the above analysis, is $\lambda_{\max}=S+1/S$. For
$\mu > 4$ and large but finite $N$, we therefore expect that the
distribution of the largest eigenvalue of ${\bf M}$ is Tracy-Widom,
but with a power-law tail of index $\mu$ that very slowly disappears
when $N \to \infty$. Our numerical results are in full agreement with
this expectation (see Fig. 1). When $\mu < 4$, on the other hand, the
number of large entries increases with $N$. However, when $\mu$ is
larger than $2$, such as to insure that the eigenvalue spectrum still
converges to the Wigner semi-circle, the number of row or columns
where two such large entries appear still tends to zero, as
$N^{2-\mu}$. Therefore, the above analysis still holds: for each
large element $S_{ij}$ exceeding unity, one eigenvalue
$\lambda=S_{ij}+S_{ij}^{-1}$ will pop out of the Wigner sea.  Even if
the eigenvalue {\it density} tends to zero outside of the interval
$[-2,2]$ when $2 < \mu < 4$, the {\it number} of eigenvalues exceeding
$2$ (in absolute value) grows as $N^{2-\mu/2} \ll N$. The $k$ largest
entries are well known to be given by a Poisson point process with
Fr\'echet intensity \cite{Sosh2}; the order of magnitude of the $k$th largest entry
is $A N^{2/\mu-1/2}/k^{1/\mu}$ which diverges with $N$, such that in
this regime the eigenvectors become strictly localized ($v^*=\pm
1/\sqrt{2}$). The largest eigenvalues are then equal to the largest
entries and are themselves given by a Poisson point process with
Fr\'echet intensity, as proven by Soshnikov in the case $\mu < 2$
\cite{Sosh2}.  His result therefore holds in the whole range $\mu <
4$. Finally, the marginal case $\mu=4$ is easy to understand from the
above discussion. The number of entries exceeding one remains of order
unity as $N \to \infty$; the distribution of the largest entry $S$ is
Fr\'echet with $N$-independent parameters: \be P_{\mu=4}(|S|)=\frac{4
A^4}{|S|^{5}} \exp\left[-\frac{A^4}{|S|^{4}}\right].  \ee
\begin{figure}
\begin{center}
\psfig{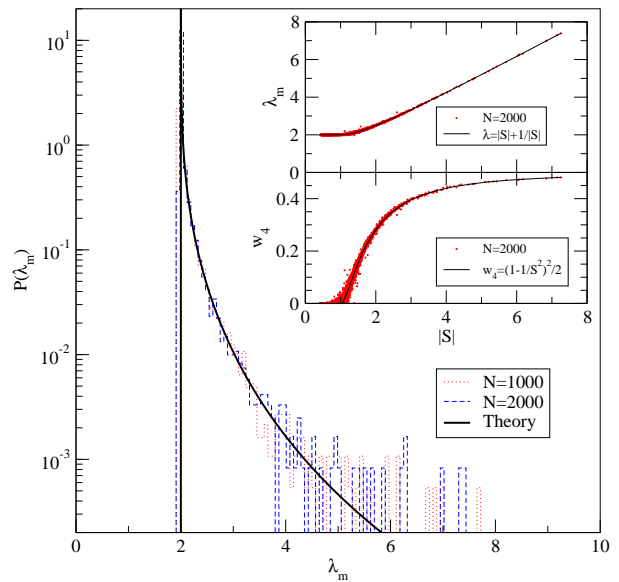} 
\end{center}
\caption{Histogram of $\lambda_{\max}$ for $\mu=4$ for matrices of
size 1000 and 2000. The solid line shows the transformed Fr\'echet
distribution with the tail amplitude $A$ set to the value used in
simulation. Note that at $\lambda=2$, there should be a Dirac delta
which is impossible to distinguish at this scale from the integrable
singularity at $\lambda=2^+$. Top inset: Scatter plot of largest eigenvalue
($\lambda_{\max}$) {\em vs} largest absolute element ($|S|$).
Theory predicts $\lambda_{\max}=2$ with $N^{-1/3}$ fluctuations
for $|S|<1$ and $\lambda_{\max}=|S|+1/|S|$ with $N^{-1/2}$ fluctuation
for $|S|>1$. Bottom inset: Scatter plot of the inverse participation
ratio ($w_4$) of the top eigenvector {\em vs} $|S|$ compared to the
prediction from Eq.\ (\ref{vstar}) (bottom inset). Similar scatter
plots were obtained for other values of $\mu$.}
\label{Fig2}
\end{figure}

The probability that $|S|$ exceeds $1$ is $\varphi=1-e^{-A^4}$, in which case $\lambda_{\max}=|S|+|S|^{-1}$; 
otherwise, with probability $1-\varphi$, $\lambda_{\max}=2$. This characterizes entirely the asymptotic distribution 
of the largest eigenvalue in the marginal case $\mu=4$: it is a mixture of a $\delta$-peak at $2$ and a
transformed Fr\'echet distribution. Note that this asymptotic distribution is non-universal since it depends 
explicitly on the tail amplitude $A$. Our argument also predicts the structure of the eigenvector when $|S|$ is
finite, see Eq. (\ref{vstar}). Again, all these results are convincingly borne out by numerical simulations, see Fig.\ 2. The statistics of the second, third, etc. eigenvalues could be understood along the same lines.

We now turn to the case of sample covariance matrices, important in many different contexts. The `benchmark' spectrum of
sample covariance matrix for {\sc iid} Gaussian random variables is well known, and given by the Mar\v{c}enko-Pastur distribution
\cite{MP}. Here again, the spectrum has a well defined upper edge, and the distribution of the largest eigenvalue is 
Tracy-Widom (see e.g. \cite{BBAP}). What happens if the random variables have heavy tails? More precisely, we consider 
$N$ times series of length $T$ each, denoted $x_i^t$, where $i=1,..,N$ and $t=1,...,T$. The $x_i^t$ have zero mean and 
unit variance, but may have power-law tails with exponent $\mu$. For example, daily stock returns are believed to have
heavy tails with an exponent $\mu$ in the range $3-5$ \cite{book}. The empirical covariance matrix $\bf C$ is defined as:
\be
C_{ij} = \frac{1}{T} \sum_t x_i^t x_j^t.
\ee
When the time series are independent, and for $T$ and $N$ both diverging with a fixed ratio $Q=T/N$, 
the eigenvalues of $\bf C$ are distributed in the interval $[(1-Q^{-1/2})^2,(1+Q^{-1/2})^2]$. 
When $T \to \infty$ at fixed $N$, all eigenvalues tend to unity, as they should since the 
empirical covariance matrix converges to its theoretical value, the identity matrix. When $N$ and $T$ 
are large but finite, the largest eigenvalue of $\bf C$ is, for Gaussian returns, a distance $\sim N^{-2/3}$ away from
the Mar\v{c}enko-Pastur edge, with Tracy-Widom fluctuations. When returns are accidentally large, this may cause 
spurious apparent correlations and substantial overestimation of the largest eigenvalue of $\bf C$. Let us
be more specific and assume, as above, that one particular return, say $x_\alpha^\tau$, is exceptionaly large, 
equal to $S$. A generalisation of the above self-consistent perturbation theory, or free convolution methods, shows that
whenever $S \leq (NT)^{1/4}$, the largest eigenvalue remains stuck at $\lambda_{\max}=(1+Q^{-1/2})^2$, whereas
when $S > (NT)^{1/4}$, the largest eigenvalue becomes:
\be
\lambda_{\max}=\left(\frac{1}{Q}+\frac{S^2}{T}\right)\left(1+\frac{T}{S^2}\right);
\ee
This result again enables us to understand the statistics of $\lambda_{\max}$ as a function of the tail exponent $\mu$.
For $N$ times series of {\sc iid} random variables, of length $T$ each, the largest element is of order $(NT)^{1/\mu}$.
For $\mu > 4$, this is much smaller than $(NT)^{1/4}$ and, exactly as above, we expect the largest eigenvalue of $\bf C$
to be Tracy-Widom, with possibly large finite size corrections \cite{Burda2}. For $\mu < 4$, large `spikes' in the time series 
dominate the top eigenvalues, which are of order $\lambda_{\max} \sim N^{4/\mu-1} Q^{2/\mu-1}$ and distributed 
according to a Fr\'echet distribution of index $\mu/2$. For applications to financial data, reasonable numbers for intraday data are $\mu=3$, 
$N=500$ and $Q=2$, leading to $\lambda_{\max} \approx 8$, compared to the Mar\v{c}enko-Pastur edge located at $2.914$. This 
shows that the effect can indeed lead to anomalously large eigenvalues with no information content. In the marginal case 
$\mu=4$, as above, $\lambda_{\max}$ has a finite probability to be equal to the Mar\v{c}enko-Pastur value, and with the complementary probability it is distributed
according to a transformed Fr\'echet distribution of index $2$, with a $T$ and $N$ independent scale. The structure
of the corresponding eigenvectors can also be investigated and is again found to be partly localized when $S > (NT)^{1/4}$.
Finally, we expect similar results to hold for the Random Singular Value problem studied in \cite{Augusta}, where 
rectangular matrices corresponding to cross correlations between 
different sets of time series are considered. 

As mentioned in the introduction, the Tracy-Widom distribution for the largest eigenvalue of complex sample covariance 
matrices has deep links with the directed polymer problem in (1+1) dimension \cite{Johansson}. A naive guess would therefore be that the universality class 
of the ground state energy changes whenever the disorder of the directed polymer problem has fat tails with an exponent 
$\mu < 4$. This is in fact not correct, at least in the version of the directed polymer problem where each site 
carries a random {\sc iid} energy. In this case, simple Flory type arguments \cite{Zhang} suggest that the universality 
class in fact changes as soon as $\mu < 5$. More precisely, the energy fluctuations should scale as $N^{1/3}$ and by Tracy-Widom 
for $\mu > 5$, and as $N^{3/(2\mu-1)}$ for $2 < \mu < 5$ with a new type of limiting distribution 
(the case $\mu < 2$ corresponds to a complete stretching of the polymer and was recently solved in \cite{Hamley}). We have 
conducted new numerical simulations of this problem which indeed confirm that for $\mu > 5$, the ground state energy 
scales as $N^{1/3}$ with Tracy-Widom fluctuations, while for $\mu < 5$ the above Flory prediction seems correct. The 
distribution $P$ of ground state energy can be fitted by a geometric convolution of Fr\'echet distributions: 
$P=(1-p)(F+p F \star F+p^2 F \star F \star F +...)$, different from the pure Fr\'echet distributed reported above for 
the largest eigenvalue for $\mu < 4$. Following \cite{BBAP}, the correct mapping should in fact be onto a directed polymer 
with power-law {\it columnar} disorder. We leave this for further detailed investigations.

In summary, we have analyzed the statistics of the largest eigenvalue of heavy tailed random matrices. We have shown that
as soon as the entries have finite fourth-moment, the largest eigenvalue has Tracy-Widom fluctuations, whereas if the
fourth-moment is infinite, the largest eigenvalue diverges with the size of the matrix and has Fr\'echet fluctuations.
In the marginal case where the fourth-moment only diverges logarithmically, the distribution is a non-universal mixture 
of a delta peak and a modified Fr\'echet law. The structure of the associated eigenvector evolves from being 
completely delocalized in the Tracy-Widom case, to partially or totally localized in the Fr\'echet case. 
We have shown that similar results holds for sample covariance matrices, and that
extreme events may cause the largest empirical eigenvalue to significantly exceed the Mar\v{c}enko-Pastur edge.  

{\em Acknowledgments - } We thank Gerard Ben Arous and Sandrine P\'ech\'e for important discussions. 
GB is partially supported by the European Community's Human Potential Program
contracts HPRN-CT-2002-00307 (DYGLAGEMEM).

\end{document}